\documentclass[epj]{webofc}
\usepackage[varg]{txfonts}   
\wocname{EPJ Web of Conferences}
\woctitle{CONF12}
\begin{document}
\selectlanguage{english}
\title{Light-Meson Spectroscopy at \textsc{Compass}}
%
%

\author{Fabian Krinner\inst{1}\fnsep\thanks{\email{fabian-krinner@mytum.de}} for the \textsc{Compass} collaboration 
}

\institute{Technische Universität München - 
Physik-Department - E18
}

\abstract{%
The goal of the \textsc{Compass} experiment at CERN is to study the structure and spectroscopy of hadrons. The two-stage spectrometer has large acceptance and covers a wide kinematic range for 
charged as well as neutral particles allowing to access a wide range of reactions. Light mesons are studied with negative (mostly $\pi^-$) and positive ($p$, $\pi^+$) hadron beams with a momentum of $190\,\text{GeV}/c$.

The light-meson spectrum is measured in different final states produced in diffractive dissociation reactions with squared four-momentum transfer $t$ to the target between $0.1$ and $1.0\,(\text{GeV}/c)^2$. The flagship 
channel is the $\pi^-\pi^+\pi^-$ final state, for which \textsc{Compass} has recorded the currently world’s largest data sample. These data not only allow us to measure the properties of known resonances with high precision, 
but also to search for new states. Among these is a new axial-vector signal, the $a_1(1420)$, with unusual properties. The findings are confirmed by the analysis of the $\pi^-\pi^0\pi^0$ final state.

}
\maketitle
\section{The \textsc{Compass} experiment}
\label{sec::compass}
The \textsc{Compass} experiment is located at CERN's Prevessin area and consists of a fixed-target two-stage magnetic spectrometer. Due to its full kinematic coverage it is capable of measuring a wide range of different physics processes, 
employing secondary hadron or tertiary muon beams, impinging on various targets.
The main goals of the physics program are the study of hadron structure as well as light-meson spectroscopy, the latter of which will be the focus here.

For the analysis presented here, data taken in $2008$ are used, where a $190\,\text{GeV}/c$ negative hadron beam impinged on a li\-quid hydrogen target. This beam consisted of $97\%$ $\pi^-$ and small admixtures of $K^-$ and antiprotons. 
The particular process under study is the diffractive production of three charged pions:
\begin{equation}
\label{eq::process}
\pi^-_\text{beam}+p_\text{target} \to \pi^-\pi^+\pi^-+p_\text{recoil}
,\end{equation}
for which \textsc{Compass} has collected a total of $46\cdot10^6$ exclusive events, resulting in the world's largest data set for this process so far.

\section{Partial-Wave Decomposition}
\label{sec::massIndependent}
The main interest of the analysis is the extraction of light-meson resonances from the data, based on the assumption that the production of the three
final-state pions happens via intermediate resonances $X^-$:
\begin{equation}
\pi^-_\text{beam}+p_\text{target} \to X^-+p_\text{recoil} \to \pi^-\pi^+\pi^-+p_\text{recoil}
.\end{equation}

\subsection{The Isobar Model}

One further assumption, that is made, is the isobar model. It states that the $X^-$ do not decay directly into three pions, but instead undergo two subsequent two-particle decays with 
an additional intermediate state $\xi$ appearing, the so-called isobar. In contrast to the three-pion resonances $X^-$, the isobar states have to be well known beforehand and fixed mass shapes, e.g.
Breit-Wigner amplitudes with predetermined parameters, have to be put into the analysis. In the analysis presented here, we use the set of isobars given in table \ref{tab::isobar}.

\begin{table}
\begin{center}
\begin{tabular}{ll}
\hline
Name & $J^{PC}_\xi$  \\\hline
$f_0(500)$ & $0^{++}$ \\
$f_0(980)$ & $0^{++}$ \\
$f_0(1500)$ & $0^{++}$\\
$\rho(770$ & $1^{--}$ \\
$f_2(1270)$ & $2^{++}$\\
$\rho_3(1690)$ & $3^{--}$  \\\hline
\end{tabular}
\end{center}
\caption{Isobars used in the Partial-Wave Decomposition and their $J^{PC}_\xi$ quantum numbers.}
\label{tab::isobar}
\end{table}
With these assumptions, the complex amplitude $\psi_\text{wave}(\tau)$ depending on the five kinematical phase-space variables $\tau$ of the process is fixed by: The spin and the signs under parity and generalized charge 
conjugation $J^{PC}$ of $X^-$, its spin-projection $M$ and the naturality $\varepsilon$ of the exchange particle, the decay into an isobar $\xi$ and a pion with the relative orbital angular momentum $L$ between the two latter.
We will call the specific combinations of these properties ``partial waves'' from hereon, they are named using the following scheme:
\begin{equation}
J^{PC}M^\varepsilon\xi\pi L
.\end{equation}
With these amplitudes, we model the measured intensity distribution $\mathcal{I}(m_{3\pi},t^\prime, \tau)$ of the process in the following way, where the appearing sum runs in our case over a set of $88$ waves \cite{bigPaper}:
\begin{equation}
\mathcal{I}(m_{3\pi},t^\prime, \tau) = \left| \sum_{i~\in~\text{waves}} T_i(m_{3\pi}, t^\prime) \psi_i(\tau) \right|^2
.\end{equation}
In above formula, the parameters $T_i$ are complex-valued transition amplitudes, that determine the strengths and phases with which the individual partial waves are produced. We extract the $T_i$ from the data by performing independent
extended maximum likelihood fits in bins of the invariant mass $m_{3\pi}$ of the three-pion system and in bins of the squared four-momentum transfer $t^\prime$. We use $100$ $m_{3\pi}$ bins of $20\,\text{MeV}/c^2$ width, in the analyzed range from $0.5$ to
$2.5\,\text{GeV}/c^2$, and $11$ non-equidistant bins in $t^\prime$ in the analyzed range from $0.1$ to $1.0\,(\text{GeV}/c)^2$, which corresponds to a total of $1100$ independent fits to the data.
\subsection{Results}
The results of the partial-wave decompositions are the transition amplitudes $T_i$ for every wave $i$ and for every bin in $m_{3\pi}$ and $t^\prime$. In the following these amplitudes will be discussed in terms of intensity and relative phases. 
The intensities $\left| T_i\right|^2$ are given in number of events. Since in every bin, one global phase factor is immeasurable, only relative phases between pairs of waves carry physical meaning. They are given by:
\begin{equation}
\label{eq::phases}
\phi_{ij} = -\phi_{ji} = \mathrm{arg}(T_iT^*_j)
.\end{equation}
The results of this step are shown as the points in all following figures.

\section{Resonance-Model Fit}
\label{sec::massDependent}
\subsection{Model and Method}
Since the result of the partial-wave decomposition only gives the transition amplitudes of the waves as a function of $m_{3\pi}$ and $t^\prime$, but no information on $3\pi$ resonances and their parameters, these have to be extracted in a second step, the resonance-model fit.
To this end, we construct the spin density matrix (SDM):
\begin{equation}
\rho_{ij}(m_{3\pi},t^\prime) = T_i(m_{3\pi},t^\prime)T^*_j(m_{3\pi},t^\prime)
,\end{equation}
where the indices $i$ and $j$ denote waves. In this formulation, the immesurable phase factor in in every $m_{3\pi}$ and $t^\prime$ bin drops out. The diagonal elements of the SDM give the intensities of the single waves, while
the off-diagonal elements give their respective interferences. Since $\rho_{ij}$ is hermitian it suffices to give intensities and relative phases, defined in equation (\ref{eq::phases}), of the waves to encode the full information. Therefore, spin density matrices will 
be shown as upper triangular matrices in the following.

We parametrize the $m_{3\pi}$ dependence of the SDM elements, by modeling the transition amplitudes as:
\begin{equation}
T_i(m_{3\pi}, t^\prime) = \sum_{r\,\in\,\text{resonances}} C_{r}(t^\prime)\text{BW}_r(m_{3\pi}) + C_\text{non-resonant}(t^\prime) \text{NR}_i(m_{3\pi}, t^\prime)
,\end{equation}
where the sum runs over the respective resonance content of wave $i$. The complex-valued functions $\text{BW}_r(m_{3\pi})$ are Breit-Wigner amplitudes used to model the resonances. The real-valued functions $\text{NR}_i(m_{3\pi}, t^\prime)$ are phenomenological parameterizations for the non-resonant contribution in wave $i$. All these amplitudes are multiplied by complex-valued coefficients $C$, that describe strength and phase of the 
respective model component. These ``couplings'' may vary with $t^\prime$.
The fact, that the couplings, as well as the non-resonant terms may vary with $t^\prime$, while the Breit-Wigner amplitudes may not, allows to better disentangle both contributions.

For the fit presented here, we selected a subset of $14$ waves with $6$ different $J^{PC}$ quantum numbers out of the $88$ waves in the analysis model to be used in the resonance-model fit. To describe these waves, we use a total of $11$ resonances.
This model has $722$ free parameters that are determined in a $\chi^2$ fit. Most of them, however, are coupling coefficients and as enter only to the $4^\text{th}$ order in the $\chi^2$ function which makes them easy to determine. They are used to extract $t^\prime$ spectra of the resonances, which we will not cover in this article. Only $51$ parameters are ``shape parameters'', e.g. masses and width 
of the resonances, and will therefore be discussed here.
In this analysis, the SDM of $14$ waves is fitted simultaneously in all $11$ bins of $t^\prime$. Below we will present selected results from this fit.
\subsection{Results}
In the following we will discuss the fit results for the six $J^{PC}$ sectors included in the fit.
\subsubsection{$J^{PC} = 0^{-+}$ Sector}
\begin{figure}[h]
\centering
\includegraphics[width=.32\textwidth,clip]{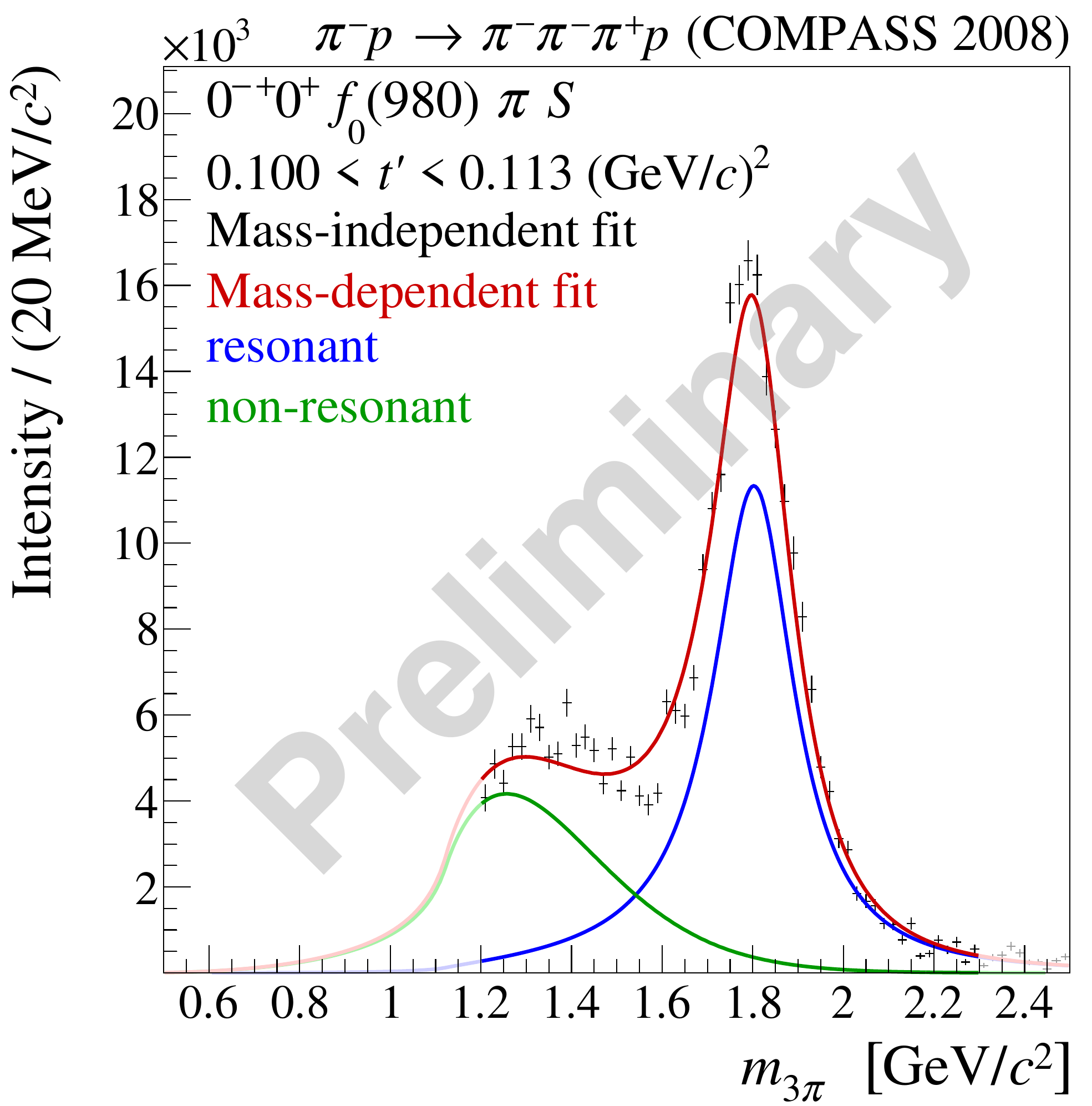}
\includegraphics[width=.32\textwidth,clip]{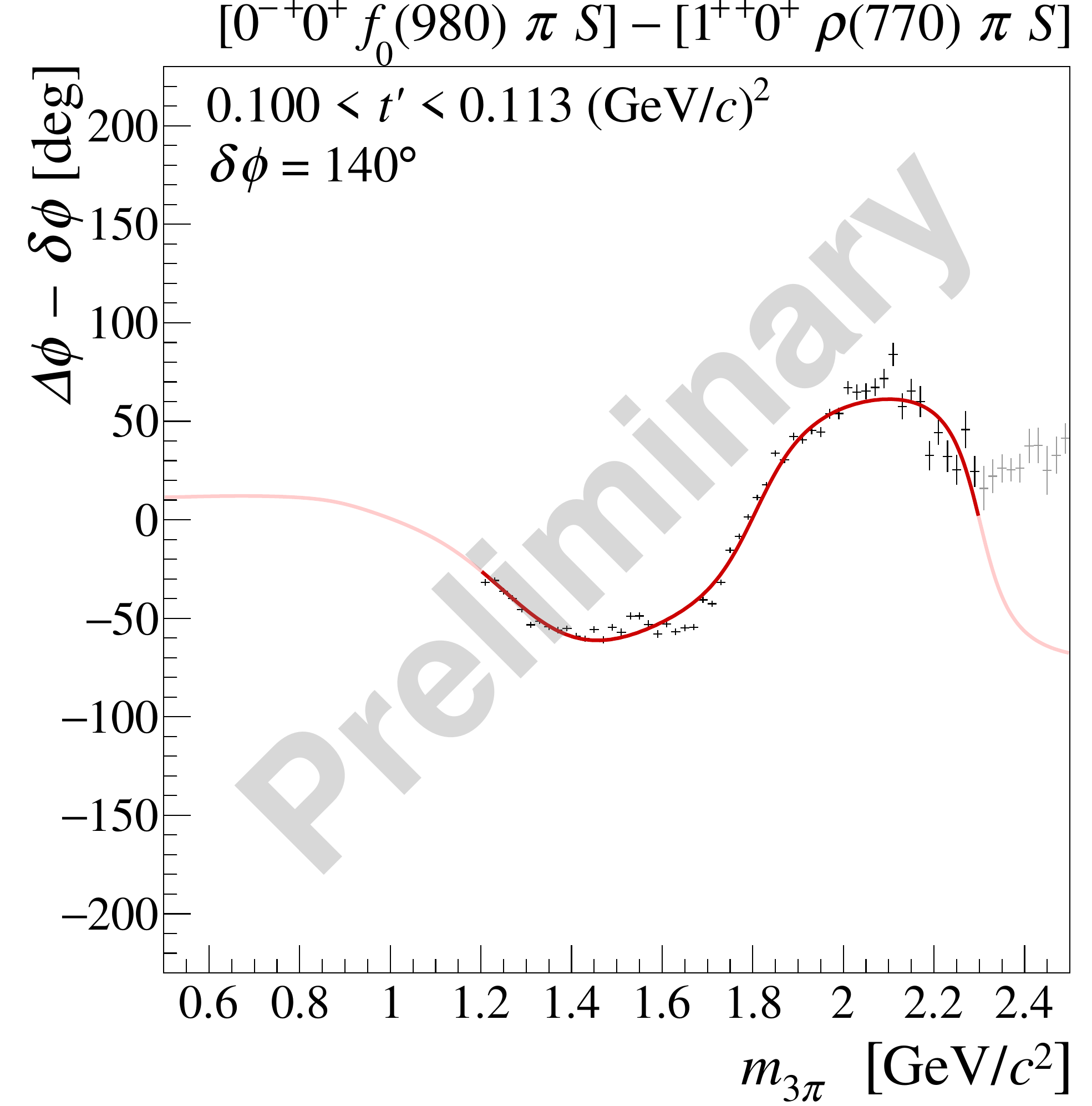}
\includegraphics[width=.32\textwidth,clip]{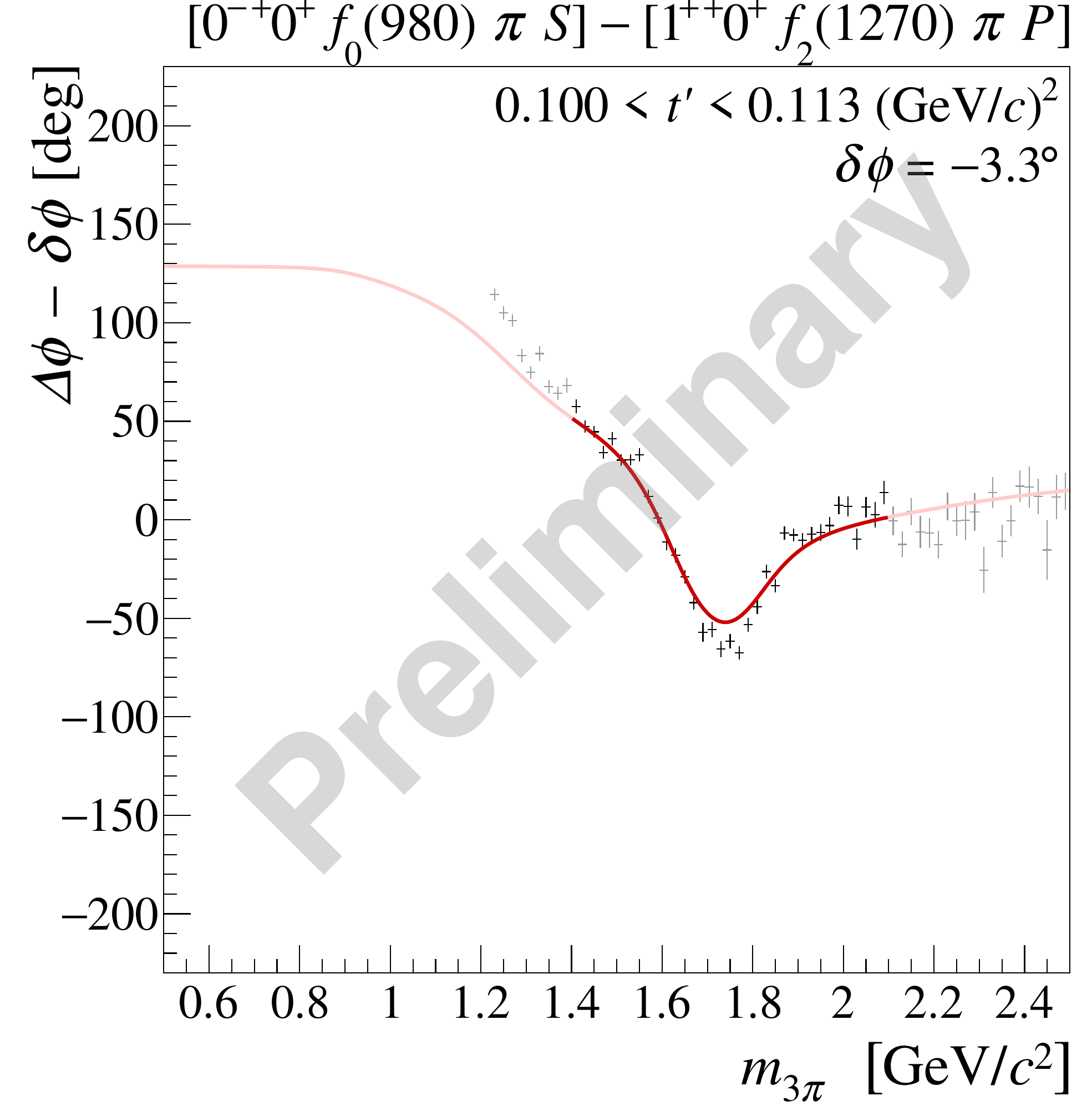}
\caption{Intensity and and phase motion for the $0^{-+}0^+f_0(980)\,\pi\,S$ wave in the lowest $t^\prime$ bin. The points show the intensity and phases extracted by the partial-wave decomposition. Black points lie within and gray ones outside the resonance-model fit-range. The red curve shows the result of the resonance model fit, while the blue and green lines in the intensity plot show only the resonant and non-resonant parts of the model. Less saturized colors give the extrapolations outside the fit range.}
\label{fig::zeroMP}       
\end{figure}

For the sector with $J^{PC} = 0^{-+}$, only the $0^{-+}0^+f_0(980)\,\pi\,S$ wave was included. In this wave a clear peak corresponding to the well-known $\pi(1800)$ resonance is seen (see figure \ref{fig::zeroMP}). In addition, rapid phase motions w.r.t. other waves are observed in the $1.8\,\text{GeV}/c^2$ mass region. The $\pi(1800)$ resonance parameters are extracted with small uncertainties, as shown in table \ref{tab::params}.

\subsubsection{$J^{PC} = 1^{++}$ Sector}
\begin{figure}[h]
\centering
\includegraphics[width=.24\textwidth,clip]{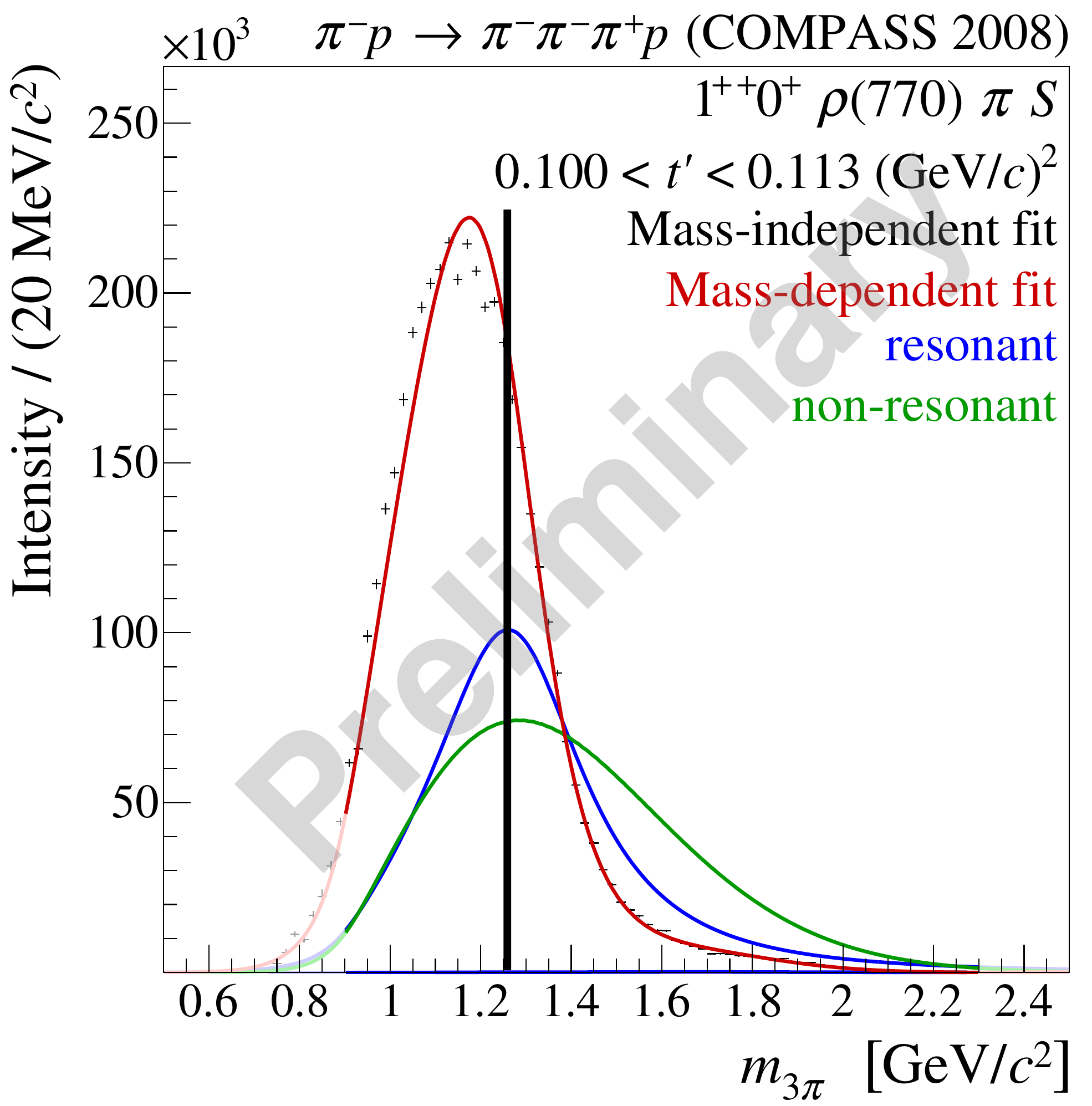}
\includegraphics[width=.24\textwidth,clip]{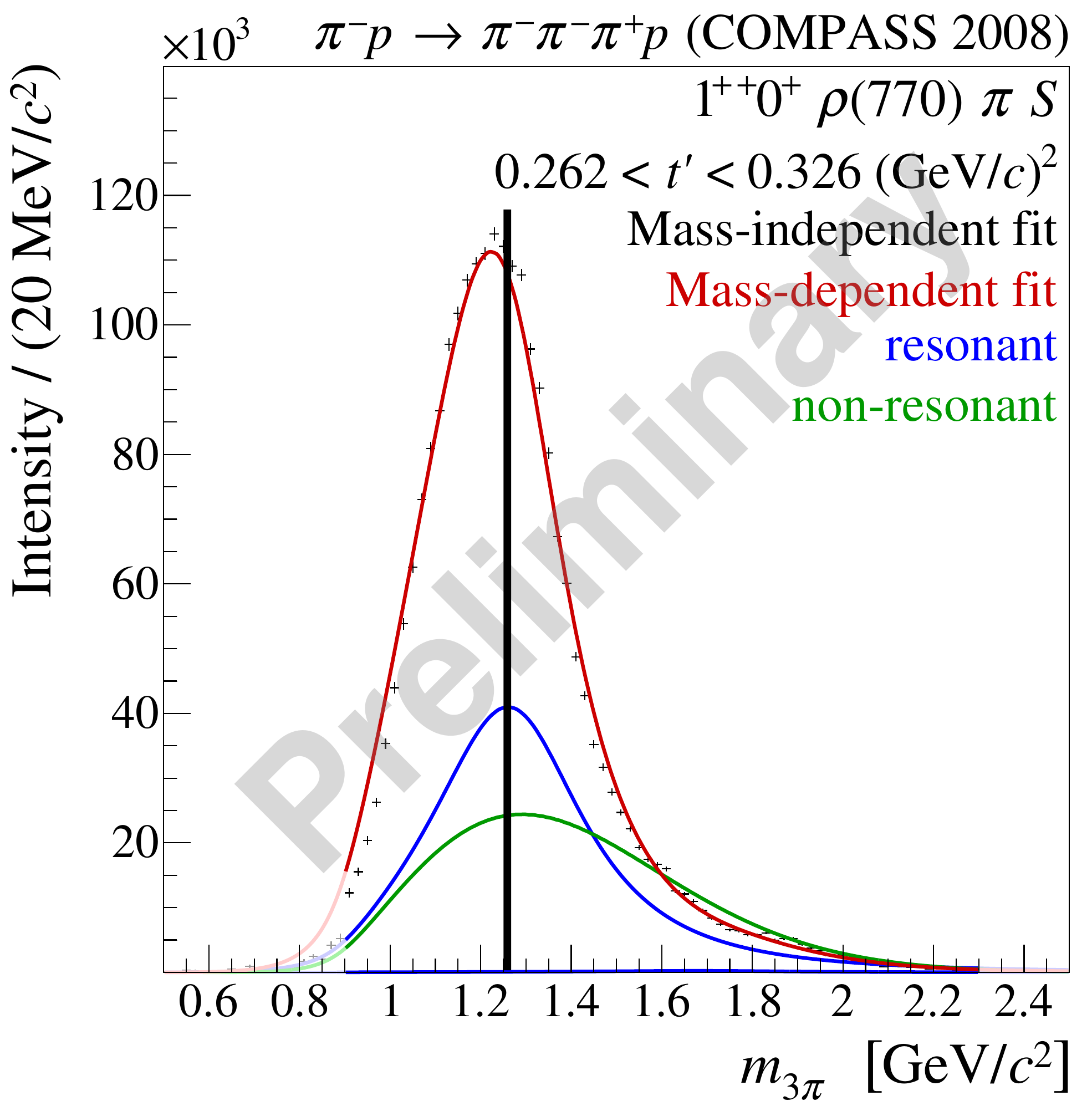}
\includegraphics[width=.24\textwidth,clip]{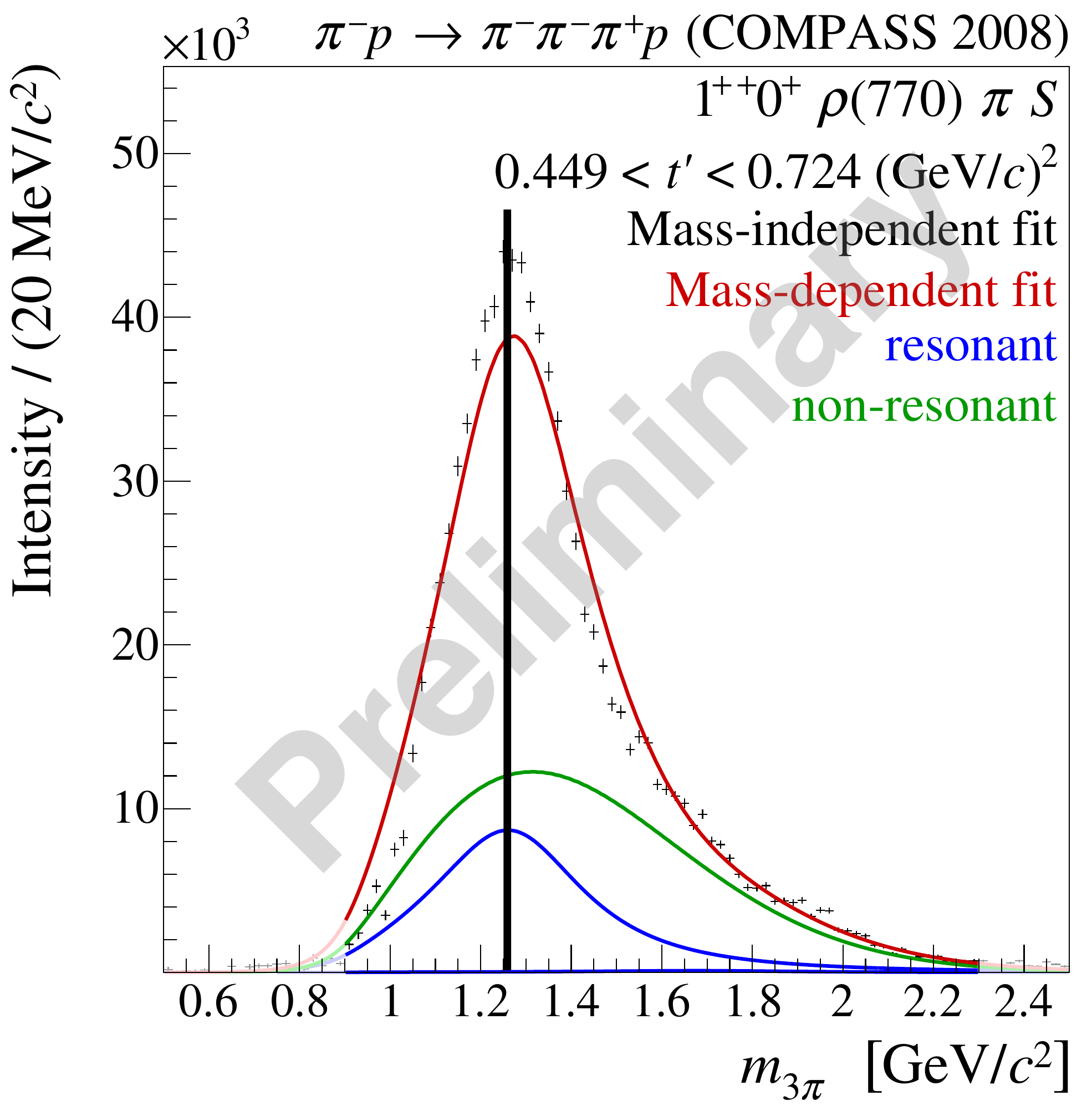}
\includegraphics[width=.24\textwidth,clip]{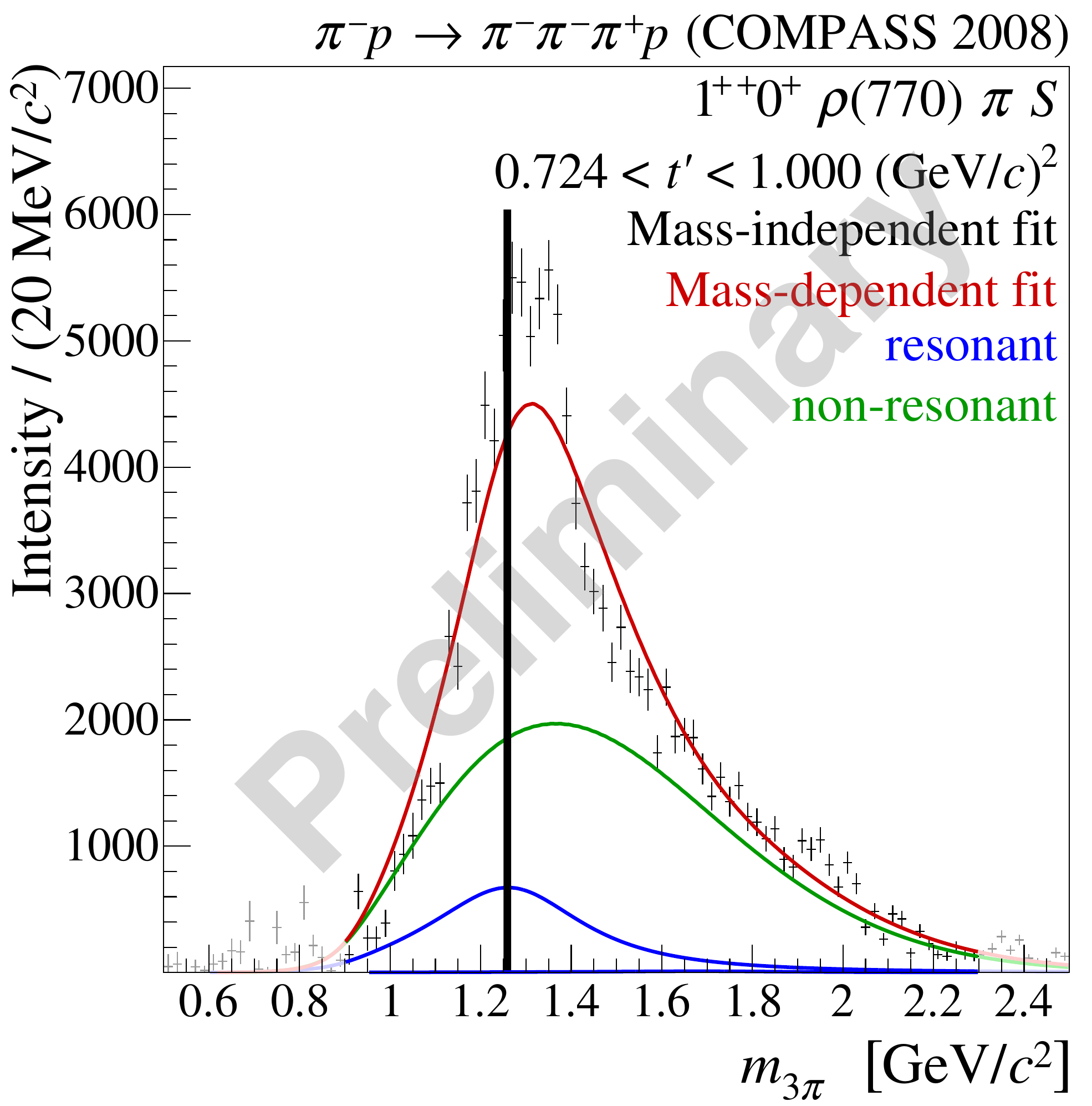}
\caption{Resonance-model fit for the $1^{++}0^+\rho(770)\,\pi\,S$ wave for four $t^\prime$ bins. The vertical line is always at the same position to illustrate the movement of the peak with $t^\prime$. This can be described by an interference between $a_1(1260)$ resonance, which does not move, and non-resonant part.}
\label{fig::onePPPeak}       
\end{figure}

Three waves with quantum numbers $J^{PC} = 1^{++}$ are included in the fit. Two of them, $1^{++}0^+\rho(770)\,\pi\,S$ and $1^{++}0^+f_2(1270)\,\pi\,P$, are described by the axial-vector resonances 
$a_1(1260)$ and $a_1(1640)$. The parameters of these resonances are extracted with reasonable systematic uncertainties from the fit (see table \ref{tab::params}). For the extraction of these resonances, the $t^\prime$ resolved analysis turned out to be important. The movement of the peak position with $t^\prime$ (see fig. \ref{fig::onePPPeak}) is explained by an interference of non-resonant and resonant contributions that changes with $t^\prime$.

The third wave in this sector is $1^{++}0^+f_0(980)\,\pi\,P$, which is consistently described by a previously unknown resonance, the $a_1(1420)$. Even though there are alternative explanations for this signal (see e.g. refs. \cite{triangle, Aceti:2016yeb, berger}) we can state, that it is compatible with a Breit-Wigner amplitude with small systematic uncertainties on the parameters (table \ref{tab::params}).

\begin{figure}[h]
\centering
\includegraphics[width=\textwidth,clip]{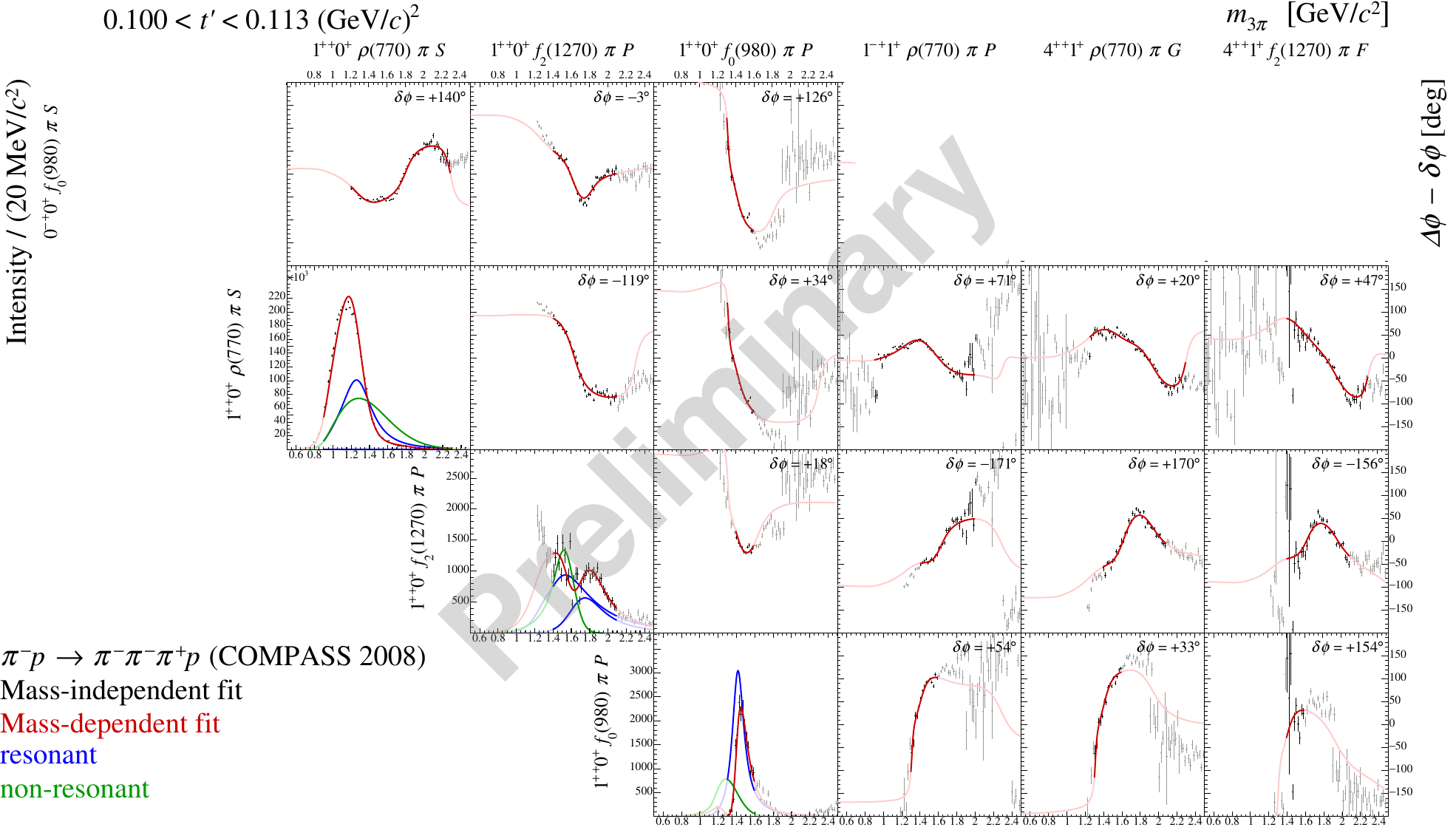}
\caption{Spin-density sub-matrix corresponding to waves with $J^{PC} = 1^{++}$ showing intensities and relative phases w.r.t. the $J^{PC} = 0^{-+}$, $1^{-+}$ and $4^{++}$ sectors.}
\label{fig::onePP}       
\end{figure}

Seletected results for this sector are depicted in figures \ref{fig::onePP} and \ref{fig::onePPPeak}.

\subsubsection{$J^{PC} = 1^{-+}$ Sector}
\begin{figure}[h]
\centering
\includegraphics[width=.32\textwidth,clip]{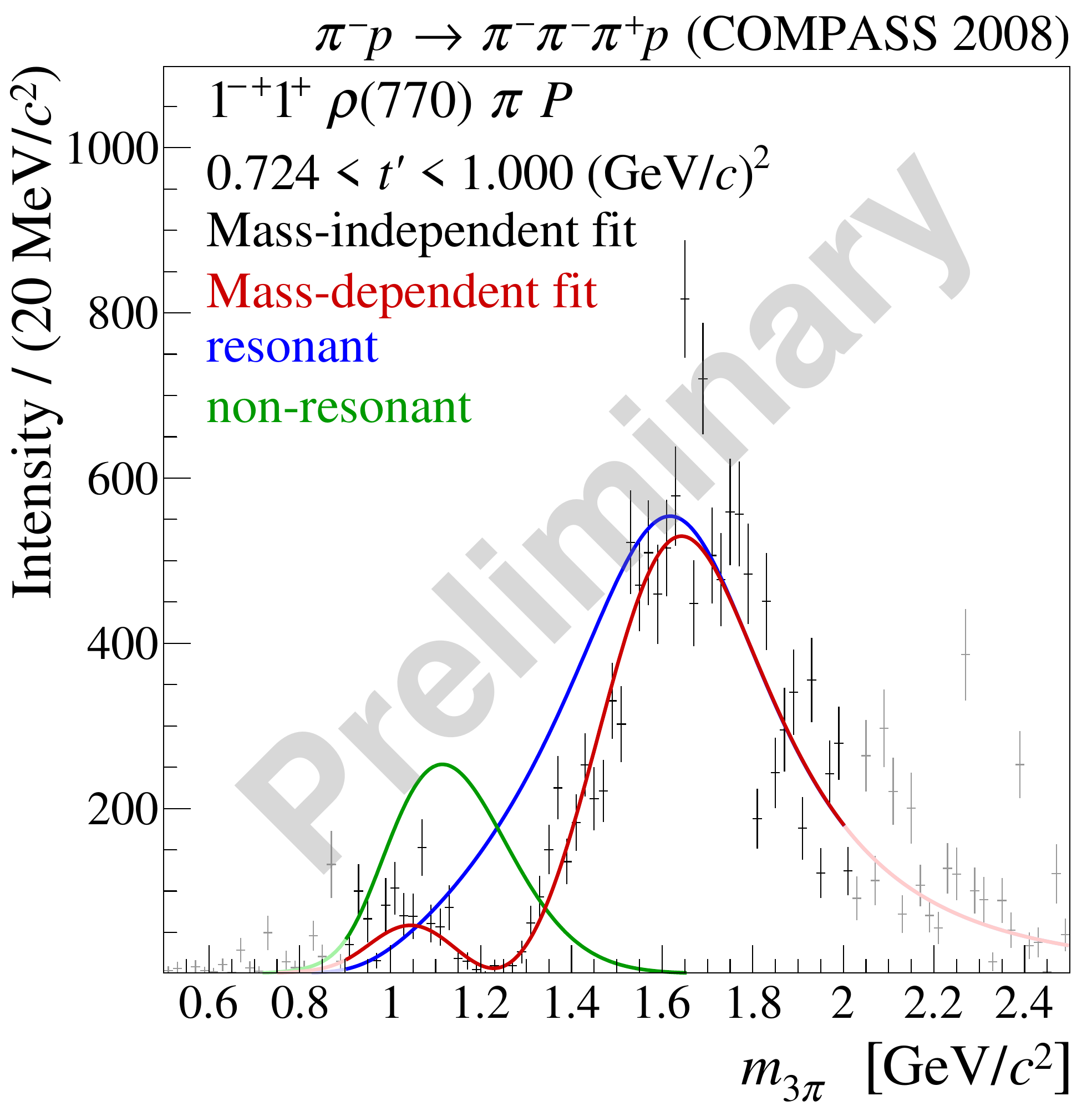}
\includegraphics[width=.32\textwidth,clip]{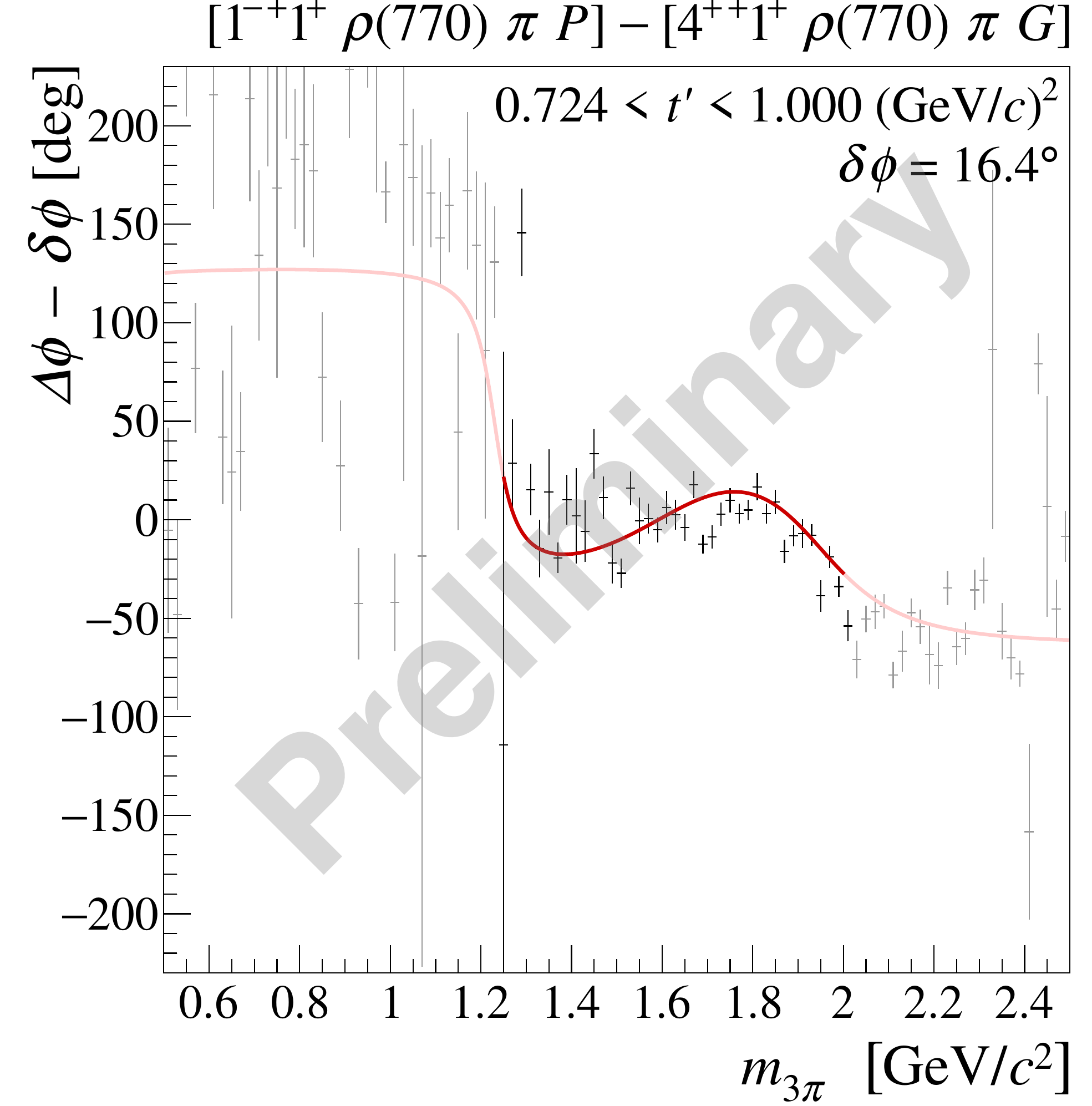}
\includegraphics[width=.32\textwidth,clip]{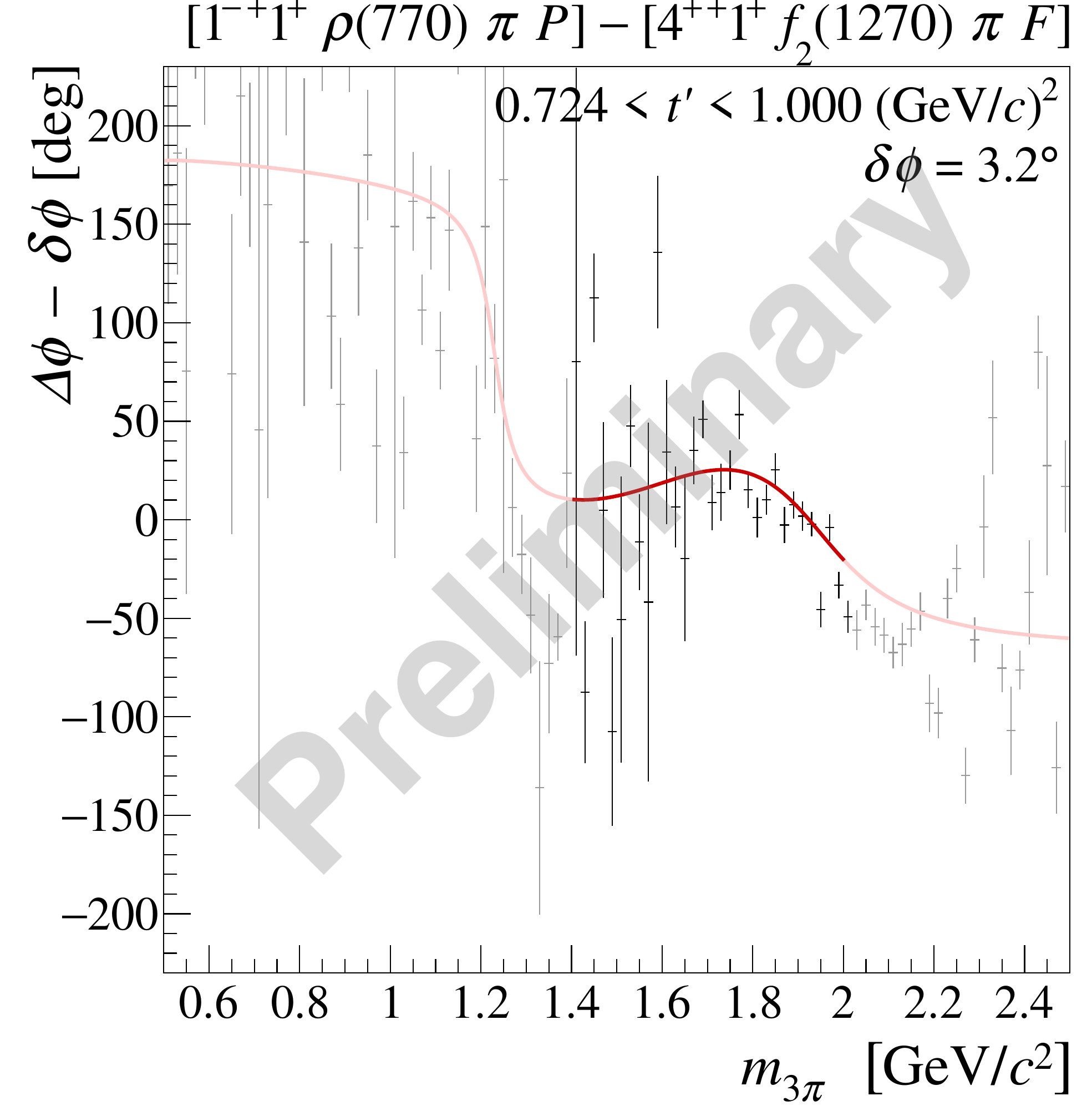}
\caption{Intensity and phase motion of the $1^{-+}1^+\rho(770)\,\pi\,P$ wave in the highest $t^\prime$ bin.}
\label{fig::oneMP}       
\end{figure}

Only one wave with $J^{PC} = 1^{-+}$ was included in the fit. Since this is a so-called ``spin-exotic'' combination of quantum numbers, a resonance in this wave could not be explained as a $q\bar q$ state. The signal extracted in our analysis is consistent with a Breit-Wigner resonance that dominates at high $t^\prime$ (see fig. \ref{fig::oneMP}), while the low-$t^\prime$ regions are dominated by the non-resonant part.
However, the width and the systematic uncertainties on the resonance parameters are rather large, thus we cannot draw a final conclusion on the existence of a resonance in this particular wave.

\subsubsection{$J^{PC} = 2^{++}$ Sector}
\begin{figure}[h]
\centering
\includegraphics[width=\textwidth,clip]{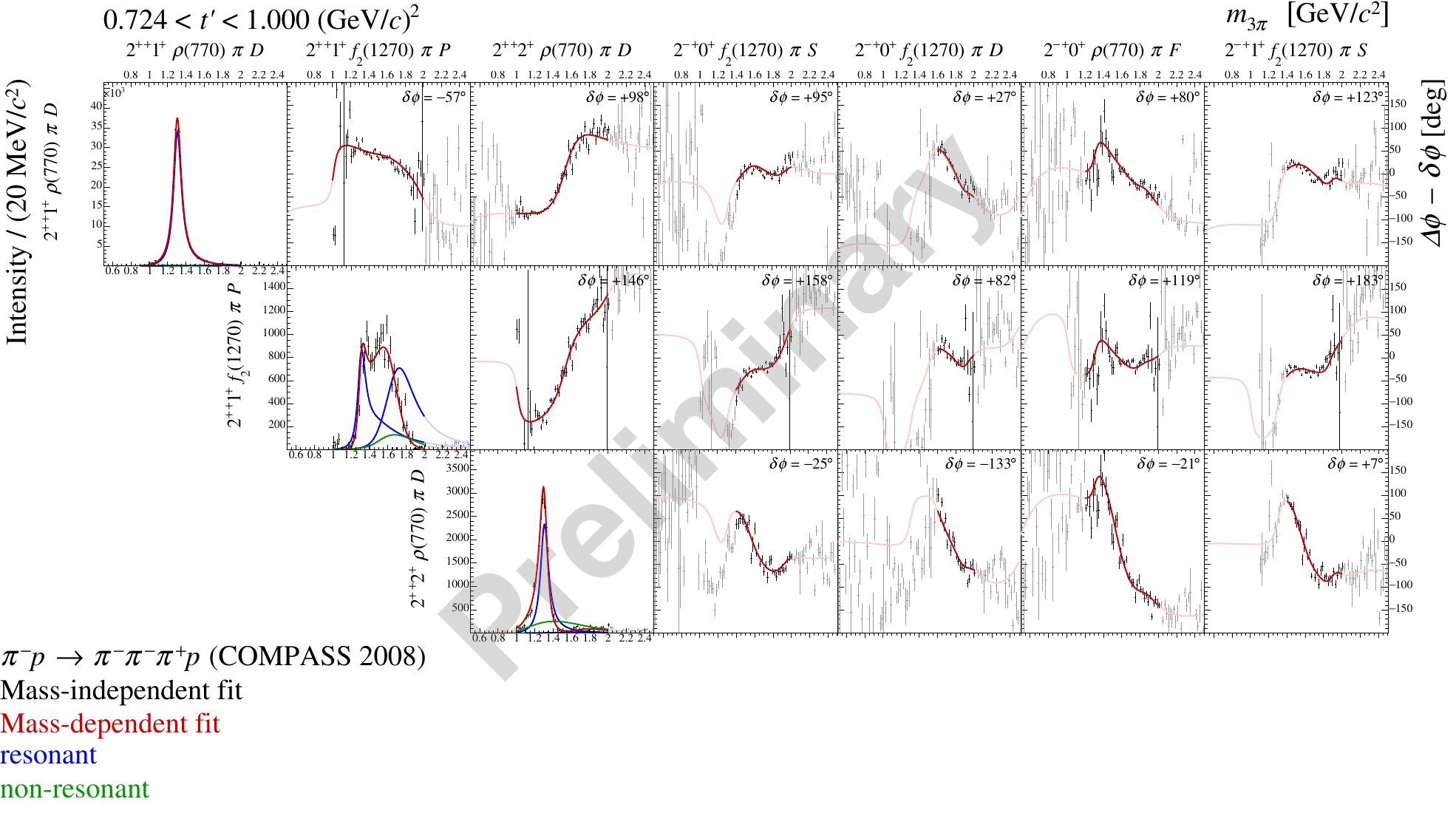}
\caption{Spin-density sub-matrix corresponding to waves with $J^{PC} = 2^{++}$. Phase differences are also shown w.r.t. the waves in the $J^{PC} = 2^{-+}$ sector.}
\label{fig::twoPP}       
\end{figure}

In the sector with $J^{PC} = 2^{++}$, three waves were included in the resonance-model fit. The waves are described by two resonances, the $a_2(1320)$ and the $a_2(1700)$. The result for this sector is shown in fig. \ref{fig::twoPP}. The peak in the $2^{++}1^+\rho(770)\,\pi\,D$ wave is also the clearest resonance signal with the smallest non-resonant contributions. Therefore, the $a_2(1320)$ is the resonance with the 
smallest systematic uncertainties in the analysis. The excited state, the $a_2(1700)$, is also determined with reasonable systematic uncertainties, tough it is produced with a much smaller intensity than the ground state.

\subsubsection{$J^{PC} = 2^{-+}$ Sector}
\begin{figure}[h]
\centering
  \includegraphics[width=.23\textwidth]{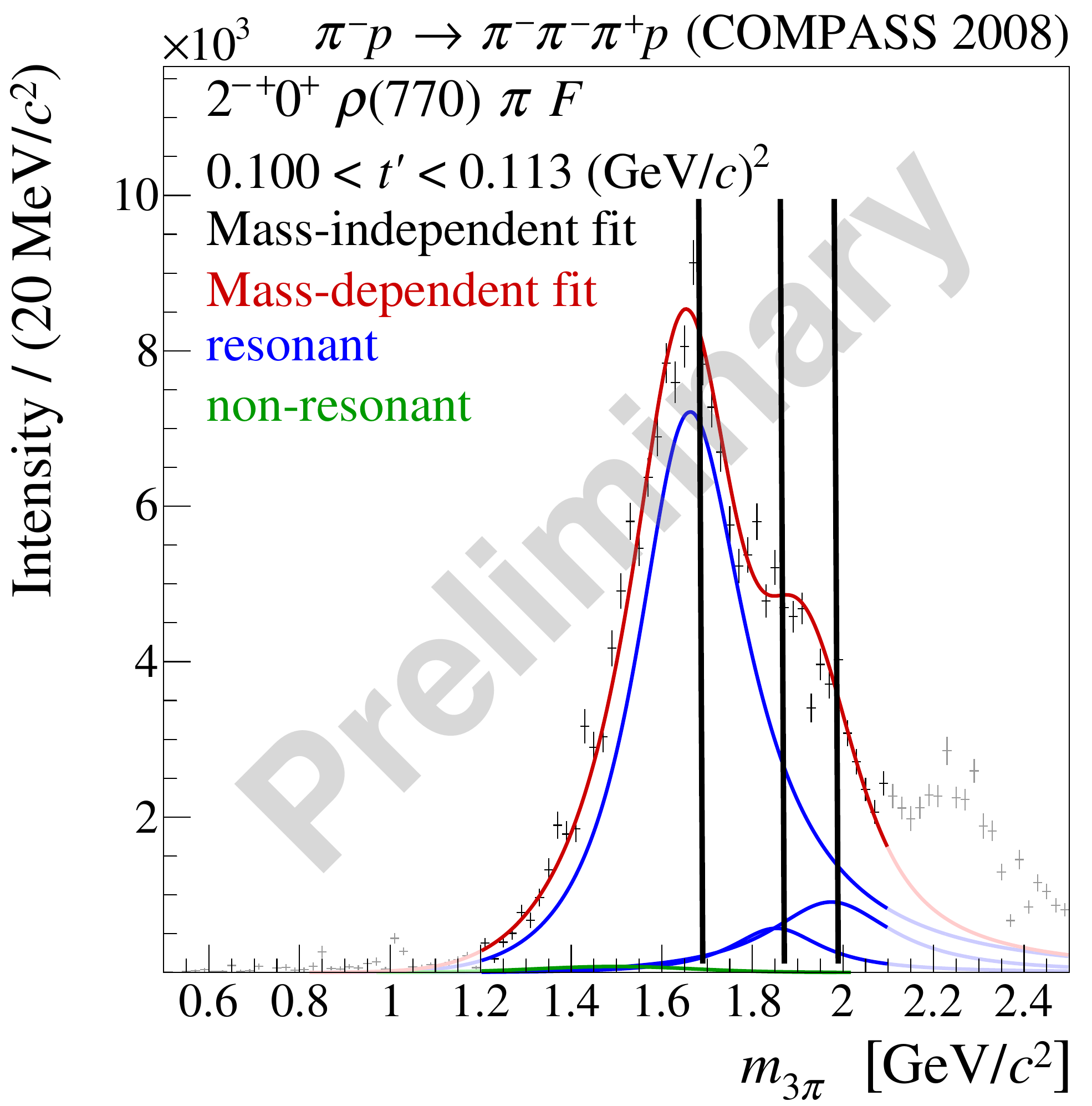}
  \includegraphics[width=.23\textwidth]{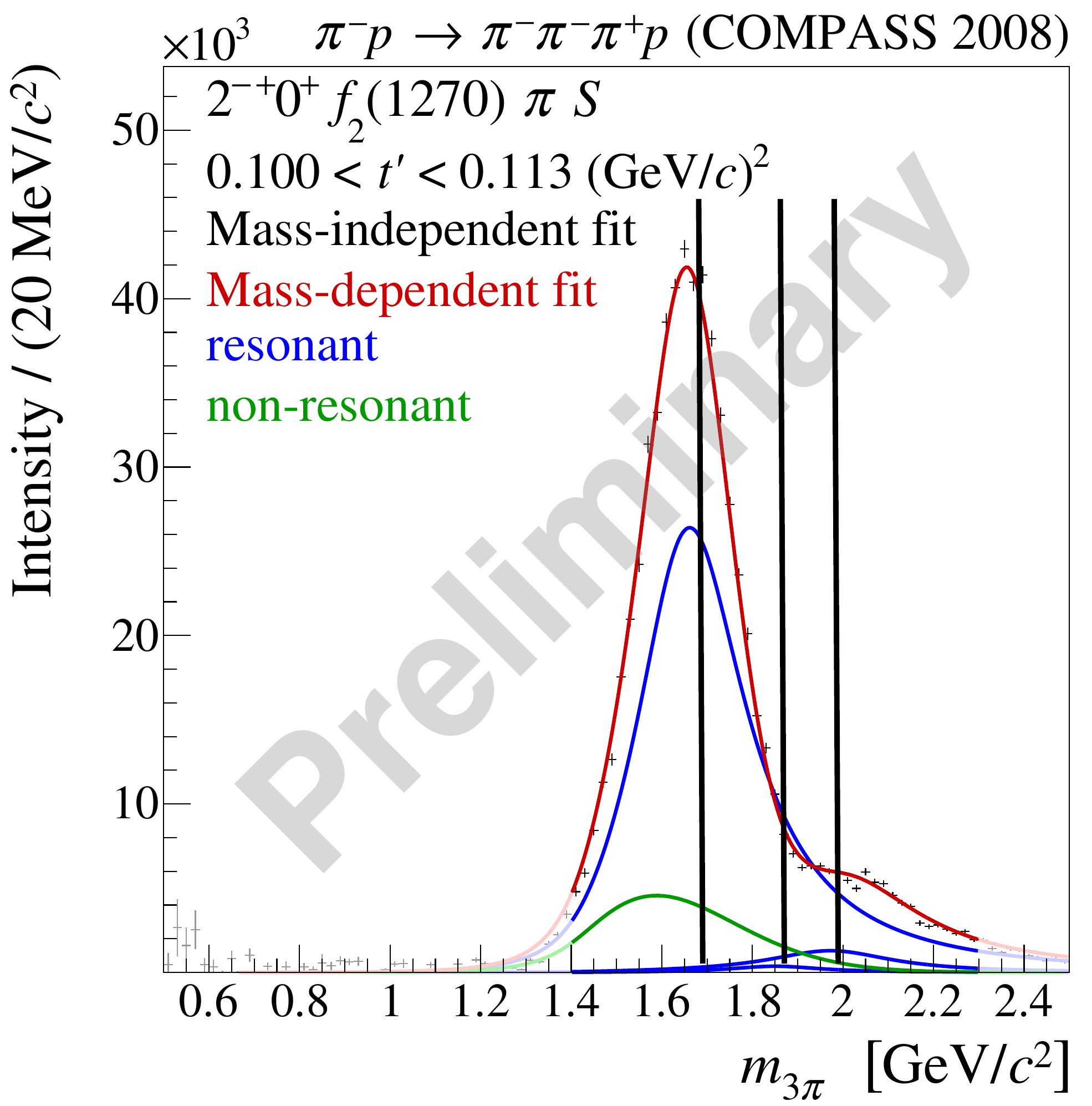}
  \includegraphics[width=.23\textwidth]{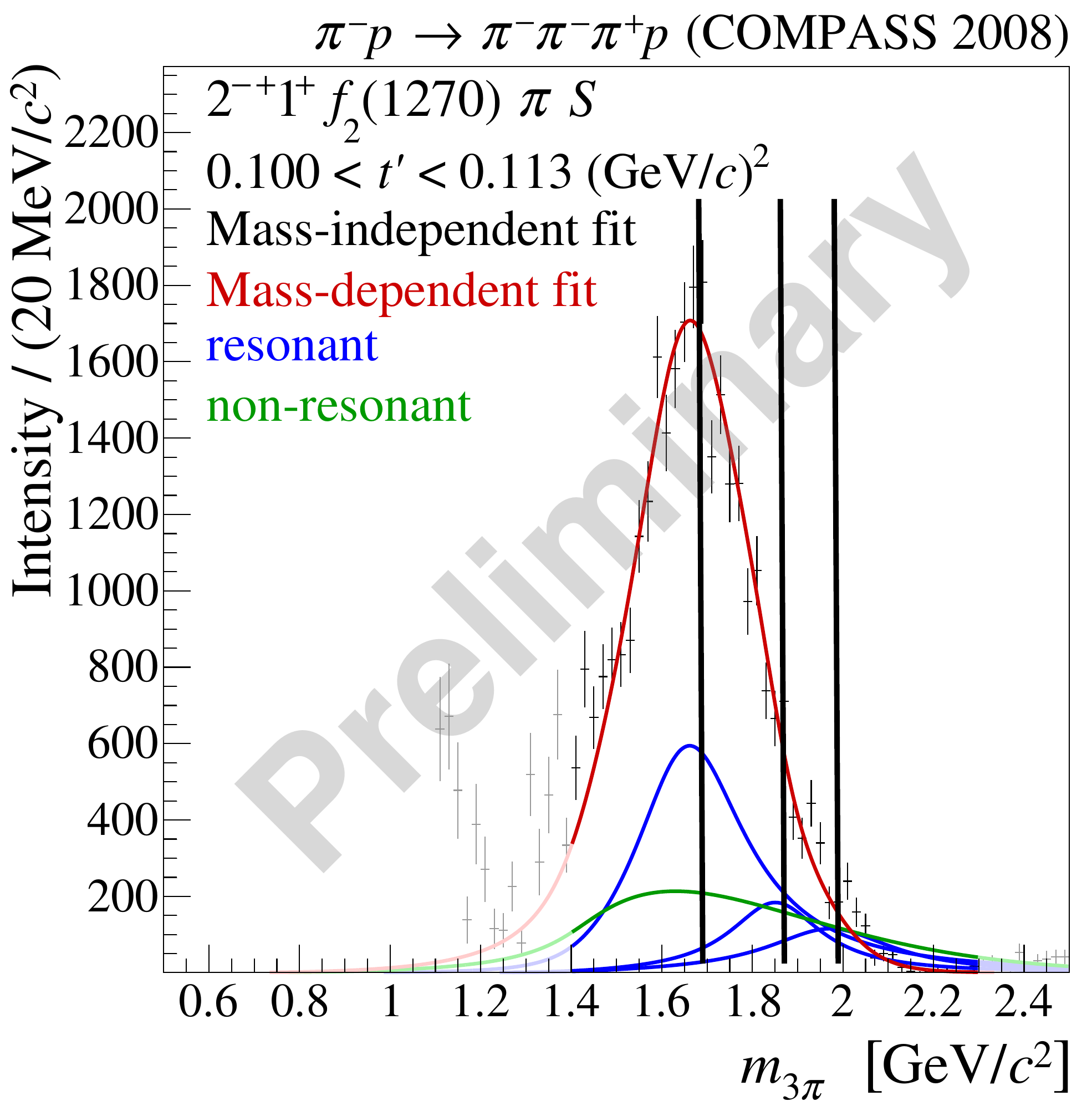}
  \includegraphics[width=.23\textwidth]{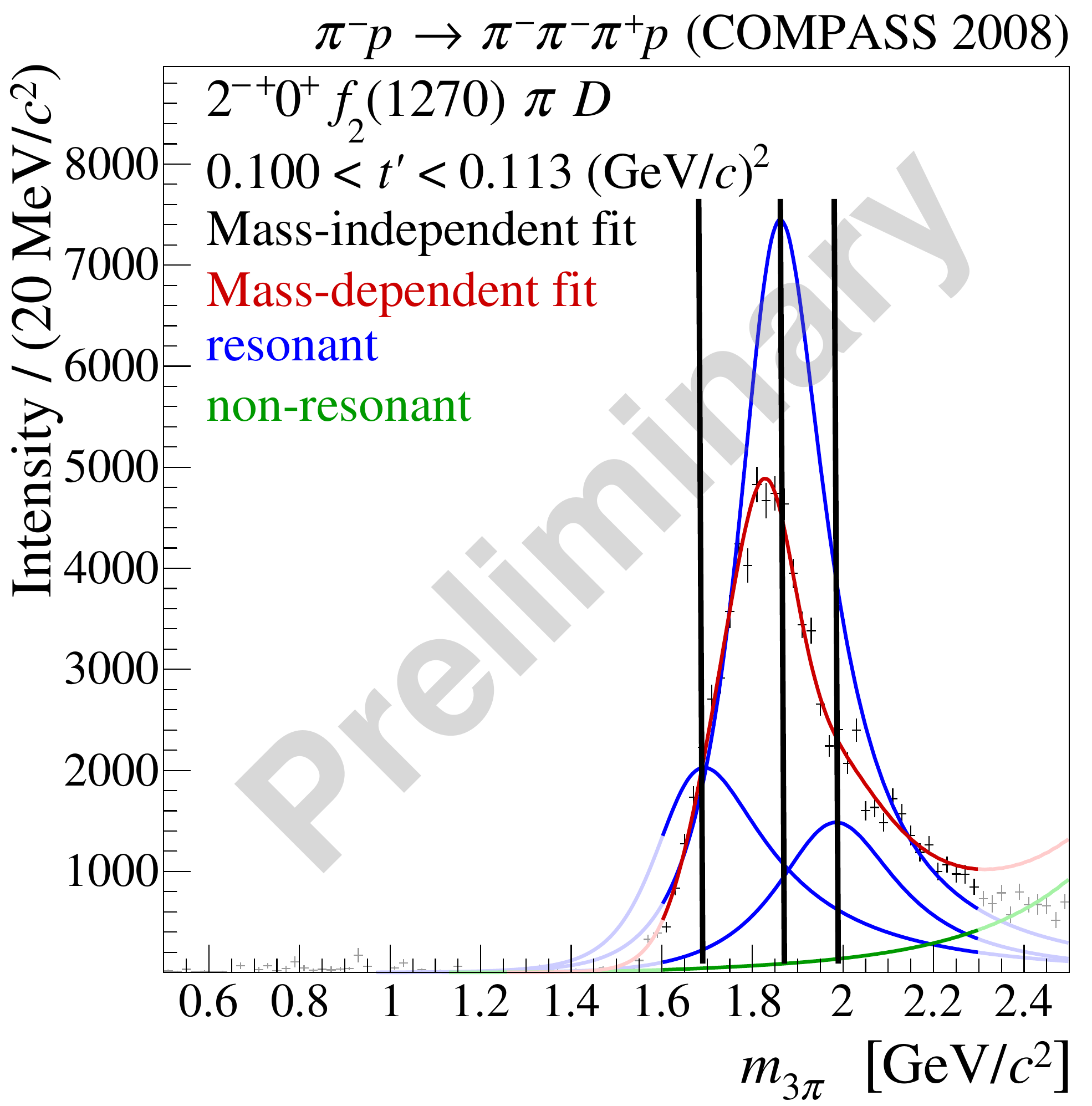}\\
  \includegraphics[width=.23\textwidth]{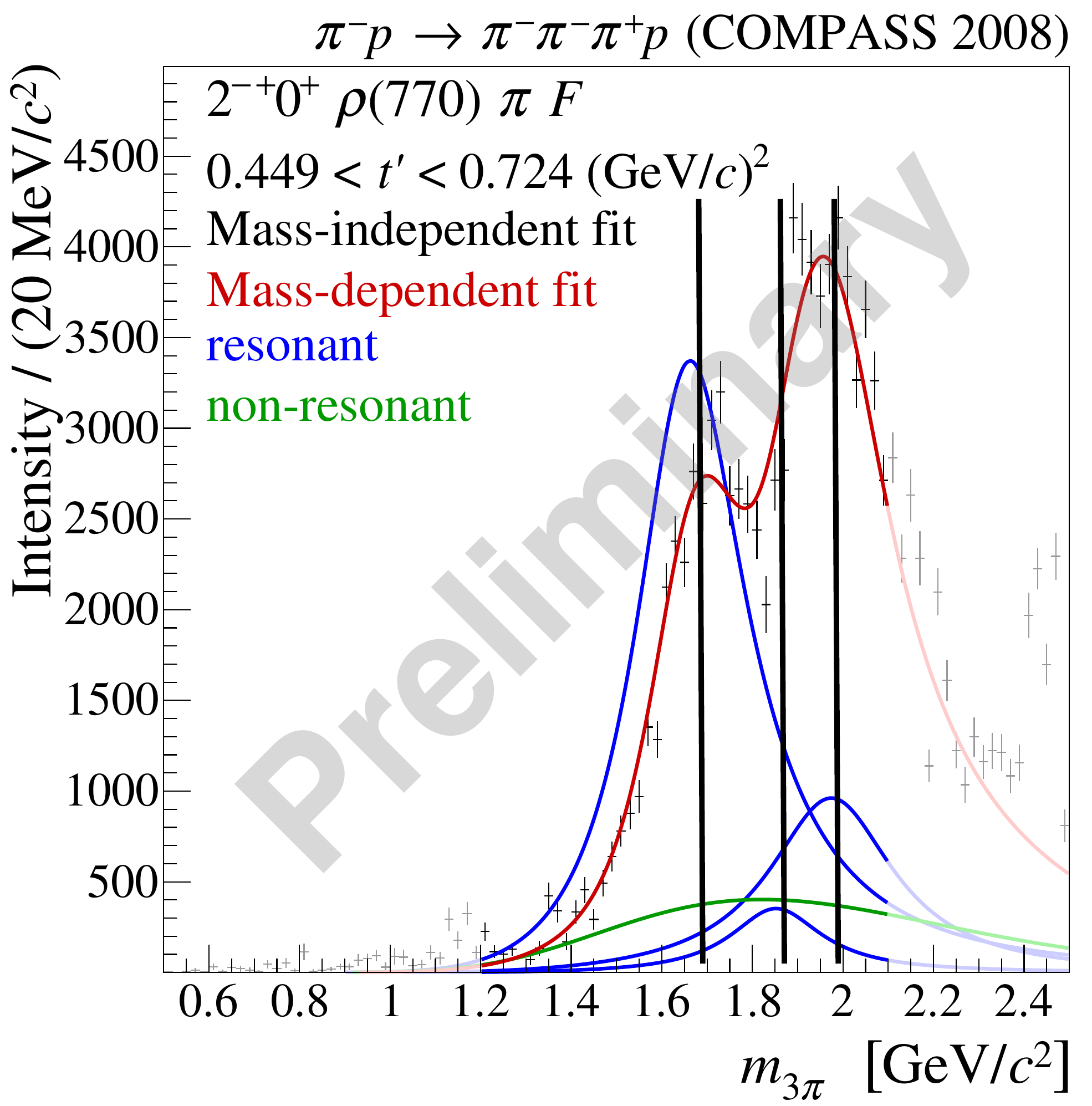}
  \includegraphics[width=.23\textwidth]{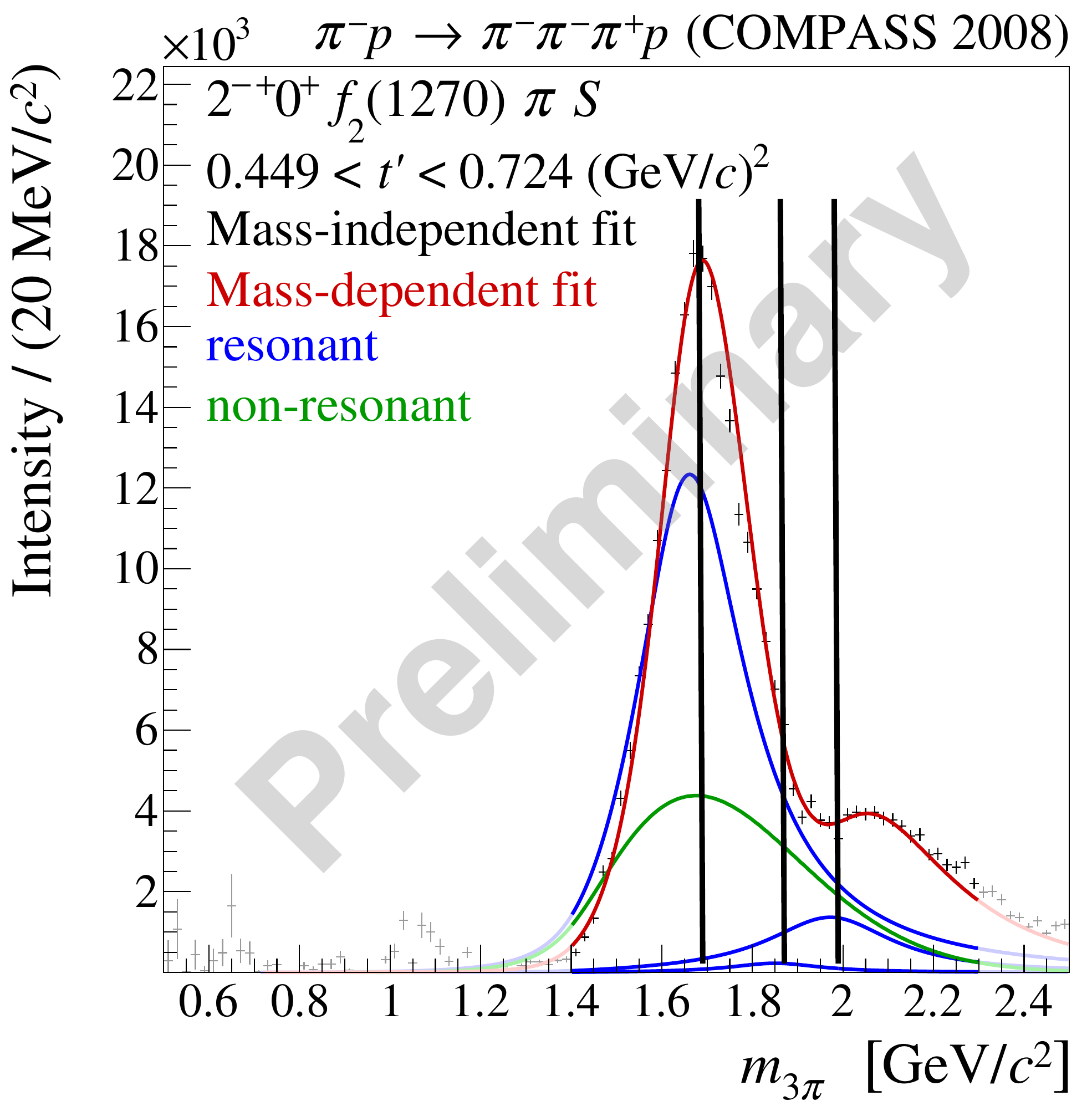}
  \includegraphics[width=.23\textwidth]{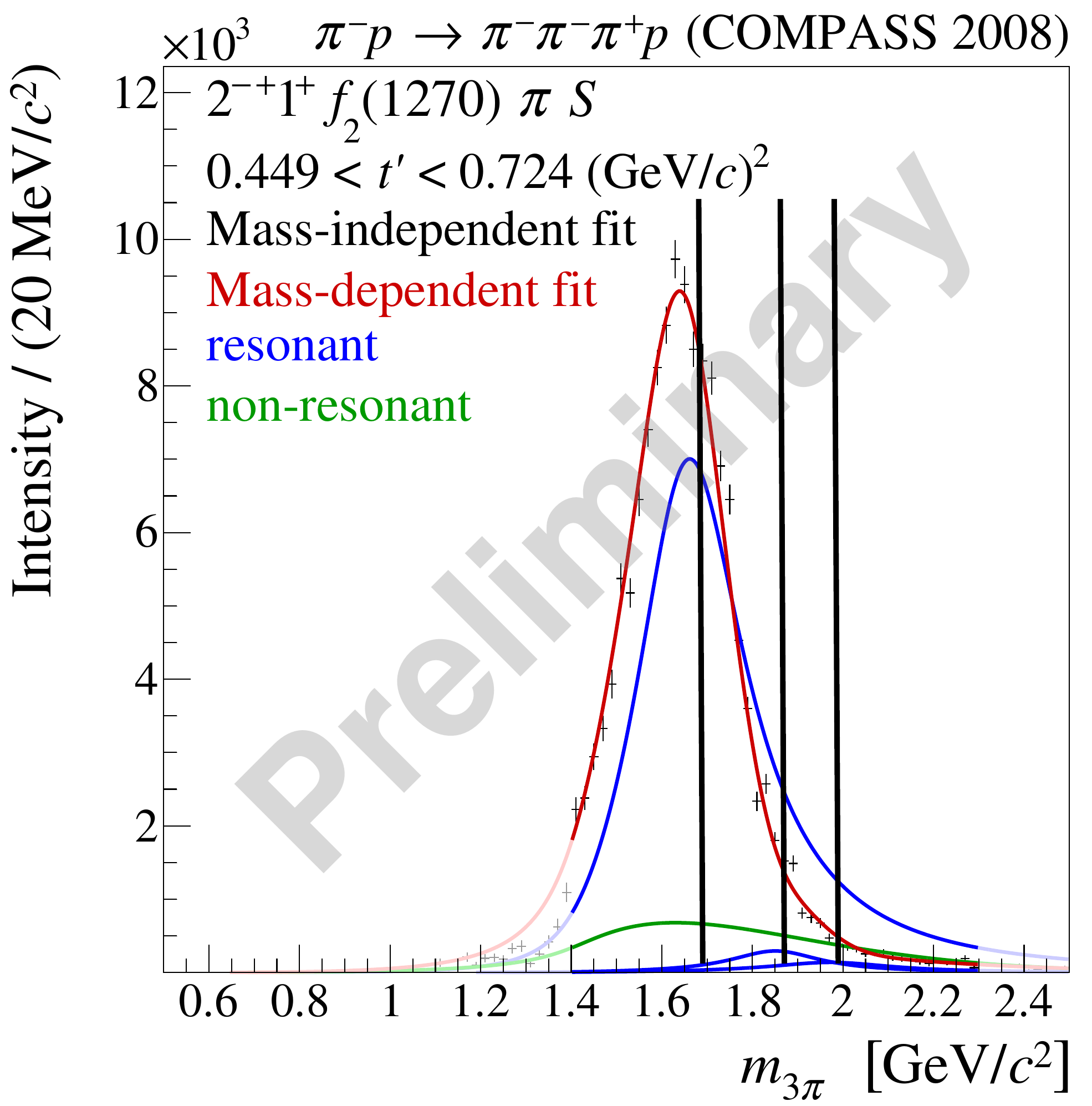} 
  \includegraphics[width=.23\textwidth]{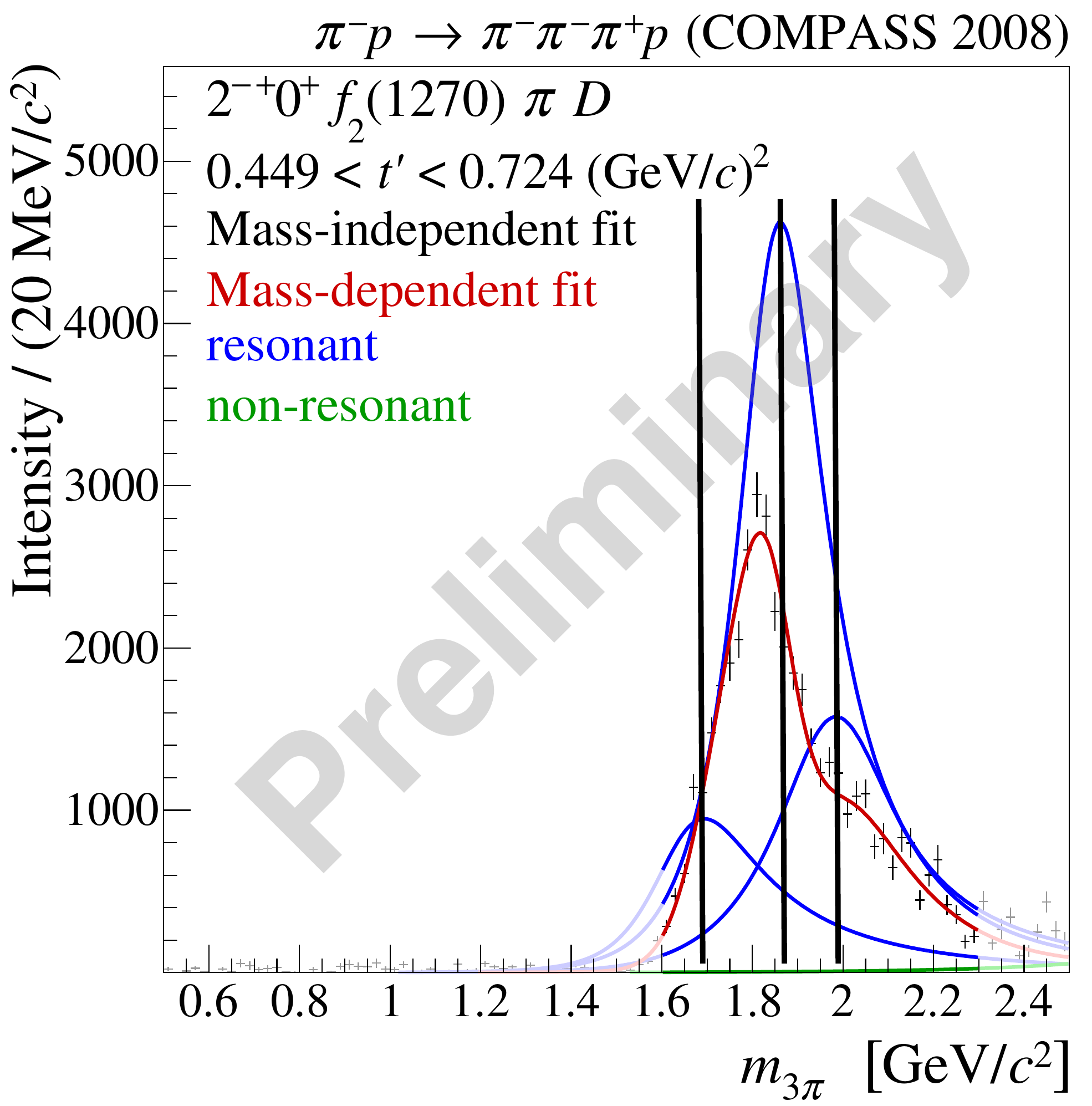}

\caption{Intensities of the four $J^{PC} = 2^{-+}$ waves in the fit, in two bins of $t^\prime$. The vertical lines show the positions of the three $\pi_2$ resonances included in the model.}
\label{fig::twoMP}       
\end{figure}

Four waves were included in the $J^{PC} = 2^{-+}$ sector, which are described using three resonances, the well known $\pi_2(1670)$ and $\pi_2(1880)$ plus one additional resonance, the $\pi_2(2005)$, which
has only been observed once before \cite{obs2005}. As can be seen in figure \ref{fig::twoMP}, the $\pi_2(1670)$ is dominant in $f_2(1270)\,\pi\,S$-wave decays, while the $\pi_2(1880)$ dominates in $f_2(1270)\,\pi\,D$-wave decays. A small amount of the $\pi_2(2005)$ is also present in every wave.

\subsubsection{$J^{PC} = 4^{++}$ Sector}
\begin{figure}[h]
\centering
  \includegraphics[width=.23\textwidth]{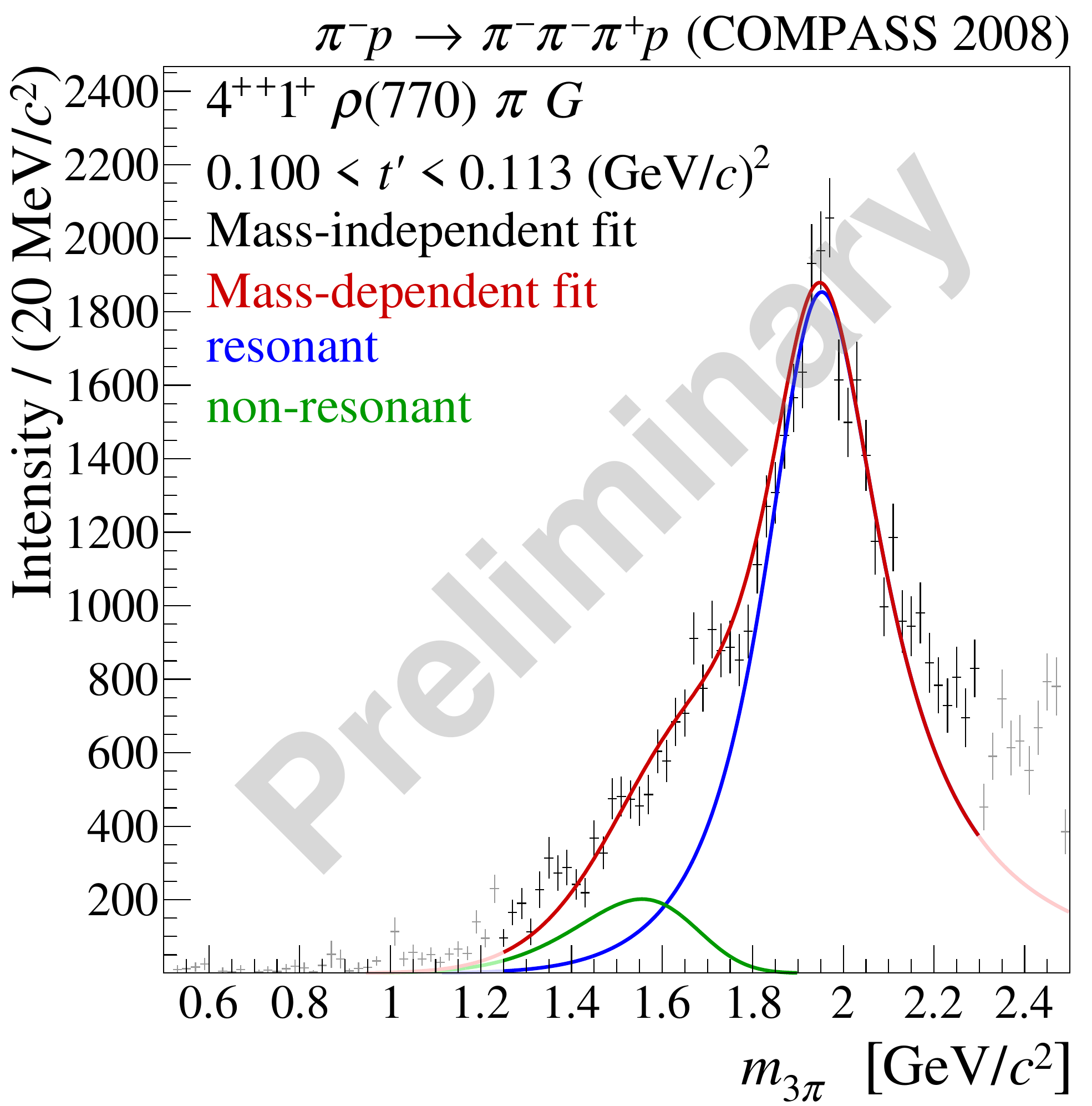}
  \includegraphics[width=.23\textwidth]{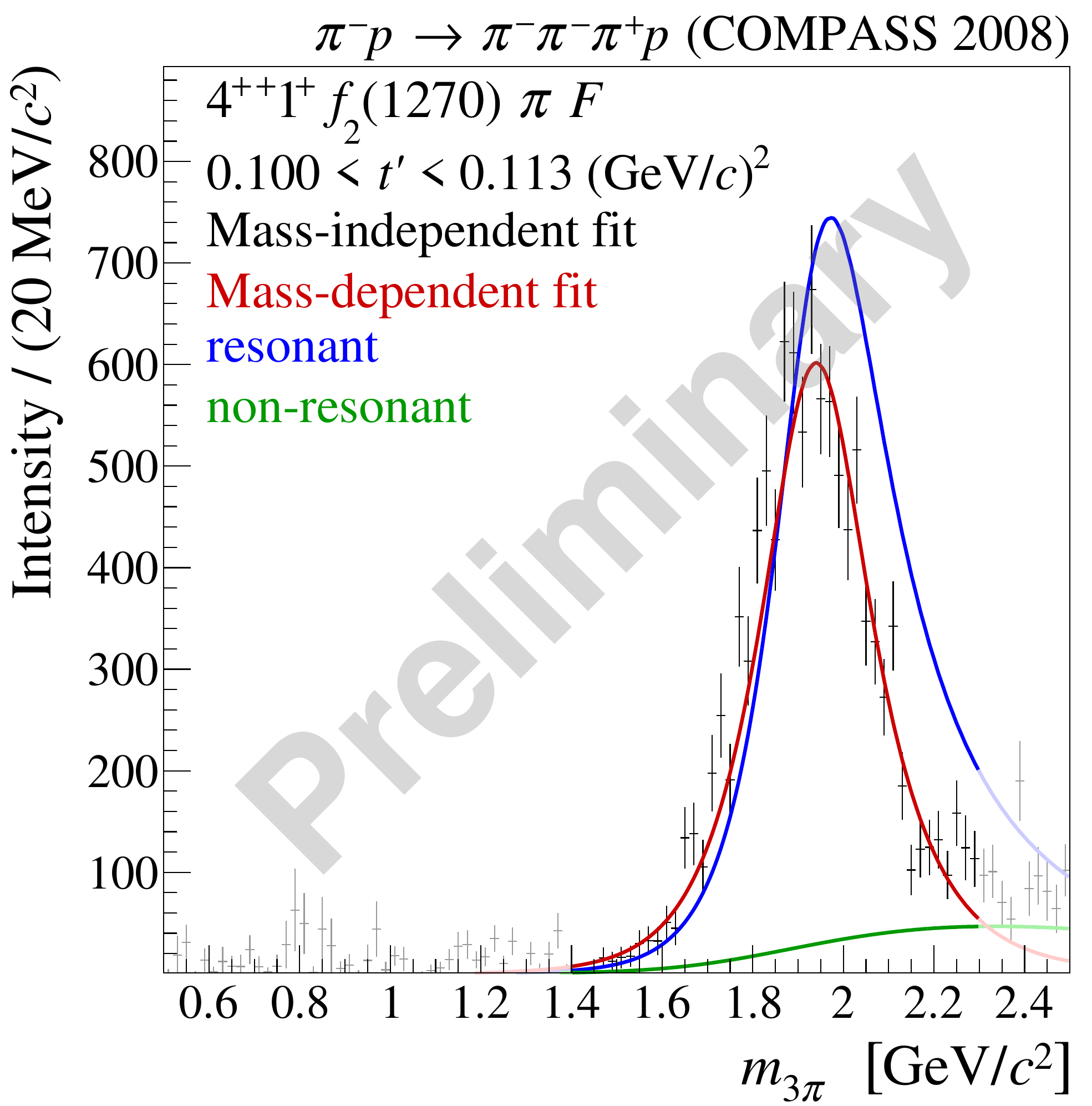}
\caption{Intensities of the two $J^{PC} = 4^{++}$ waves in the fit in the lowest $t^\prime$ bin.}
\label{fig::fourPP}       
\end{figure}

The last sector included in the analysis is the $J^{PC} = 4^{++}$ sector, for which two waves were used, shown in fig. \ref{fig::fourPP}. These are described by the $a_4(2040)$ resonance, whose mass and width is determined with 
small systematic uncertainties.
\subsection{Systematic Studies}
All results presented are part on the best fit result to the data out of $1000$ attempts with random start values and varying release orders for the parameters. These attempts were performed using a total of
$30000\,\text{CPUh}$.
To get an additional handle on the systematic uncertainties of the extracted resonance parameters, more than $200$ systematic studies with $200$ fit attempts each were performed. These studies include the variation of the:
\begin{itemize}
\item set of fitted waves,
\item resonances used to describe the waves,
\item parameterizations of the non-resonant parts,
\item fit ranges in $m_{3\pi}$,
\item binning in $t^\prime$, 
\item event selection,
\item definition of the $\chi^2$ function.
\end{itemize}
Based on the results of these studies, we are confident to have a reasonable estimate of the systematic uncertainties on the extracted resonance parameters that are summarized in Table \ref{tab::params}.
\begin{table}
\begin{center}
\renewcommand{\arraystretch}{1.2}
\begin{tabular}{lll}
\hline
Resonance 	& $m_0[\text{MeV}/c^2]$ 	& $\Gamma_0[\text{MeV}/c^2]$  \\\hline
  $a_1(1260)$ & $1298^{+13}_{-22}  $ & $403^{+0}_{-100}$  \\
  $a_1(1420)$ & $1411.8^{+1.0}_{-4.4}$ & $158^{+8}_{-8}$    \\
  $a_1(1640)$ & $1688^{+40}_{-70}  $ & $534^{+124}_{-20}$ \\
  $a_2(1320)$ & $1314.2^{+1.0}_{-3.1}$ & $106.7^{+3.5}_{-2.4}$\\
  $a_2(1700)$ & $1674^{+143}_{-32} $ & $435^{+52}_{-15}$  \\
  $a_4(2040)$ & $1933^{+13}_{-14}  $ & $334^{+22}_{-19}$  \\\hline
  $\pi(1800)$   & $1802.6^{+8}_{-3.5}   $ & $218^{+11}_{-6}$  \\
  $\pi_1(1600)$ & $1604^{+100}_{-50}  $ & $608^{+70}_{-240}$\\
  $\pi_2(1670)$ & $1644.2^{+11.5}_{-3.4}$ & $306^{+14}_{-19}$ \\
  $\pi_2(1880)$ & $1847^{+14}_{-6}    $ & $247^{+41}_{-18}$ \\
  $\pi_2(2005)$ & $1968^{+21}_{-21}   $ & $337^{+50}_{-80}$ \\\hline
\end{tabular}
\end{center}
\caption{Breit-Wigner parameters extracted from the resonance-model fit with their systematic uncertainties. Statistical uncertainties are negligible due to the large size of the data set.}
\label{tab::params}
\end{table}

\section{Conclusions and Outlook}
\label{sec::conclusion}
The large data set for the process $\pi^-_\text{beam} p_\text{target} \to \pi^-\pi^+\pi^-p_\text{recoil}$ collected with the \textsc{Compass} spectrometer allows us to preform a detailed partial-wave decomposition and 
a precise subsequent resonance-model fit.

In the performed analysis, we were able to confirm a new resonance, the $a_1(1420)$, which has been observed for the first time in a more limited analysis \cite{letter}. In addition, a signal with spin-exotic $J^{PC} = 1^{-+}$ quantum numbers was observed, which is consistent with a Breit-Wigner resonance description. Furthermore we extracted nine previously known isovector resonances.

Since the statistical uncertainties on our analysis are very small in comparison with systematic ones, good knowledge of the latter is essential. They were determined based on a large number
of systematic studies that give a good handle on their size.

The analysis of the data is not yet completed, since several questions still remain unanswered. The first concerns the level, to which the isobar model is valid. To this extent, we explore a novel analysis method that does
not use fixed parameterizations of the isobars \cite{MENUproceedings}. We also explore the role of possible non-resonant production mechanisms like the Deck effect.
The second question concerns the selection of the wave set for the partial-wave decomposition, which was done by hand up to now and thus might have introduced observer bias to the analysis. 
To study this, we are currently working on an automated model-selection procedure \cite{DRBicker}.
The last question concerns the resonance parameterizations used in the resonance-model fit. At the moment simple Breit-Wigner amplitudes and functions without phase motion for non-resonant contributions are used, but we are making efforts to use more advanced parameterizations that fulfill the requirement of analyticity and unitarity. As a first application, we study resonance pole positions in the $J^{PC} = 2^{-+}$ sector \cite{MishaThese}.


\begin{thebibliography}{}

\bibitem{bigPaper}
  C.~Adolph {\it et al.} [COMPASS Collaboration],
  arXiv:1509.00992 [hep-ex].

\bibitem{letter}
  C.~Adolph {\it et al.} [COMPASS Collaboration],
  Phys.\ Rev.\ Lett.\  {\bf 115} (2015),  082001
  doi:10.1103/PhysRevLett.115.082001.

\bibitem{triangle}
  M.~Mikhasenko, B.~Ketzer and A.~Sarantsev,
  Phys.\ Rev.\ D {\bf 91} (2015),  094015
  doi:10.1103/PhysRevD.91.094015

\bibitem{Aceti:2016yeb}
  F.~Aceti, L.~R.~Dai and E.~Oset,
  arXiv:1606.06893 [hep-ph].


\bibitem{berger}
  J.~L.~Basdevant and E.~L.~Berger,
  Phys.\ Rev.\ Lett.\  {\bf 114} (2015) 192001, 
  arXiv:1501.04643 [hep-ph].



\bibitem{obs2005}
  A.~V.~Anisovich, C.~A.~Baker, C.~J.~Batty, D.~V.~Bugg, V.~A.~Nikonov, A.~V.~Sarantsev, V.~V.~Sarantsev and B.~S.~Zou,
  Phys.\ Lett.\ B {\bf 517} (2001) 261
  doi:10.1016/S0370-2693(01)01017-6


\bibitem{MENUproceedings}
  F.~Krinner [COMPASS Collaboration],
  arXiv:1609.08514 [hep-ex].


\bibitem{DRBicker}
  K.~A.~Bicker (2016), ``Model Selection for and Partial-Wave Analysis of a Five-Pion Final State at the COMPASS Experiment at CERN'' (Doctoral Dissertation).

\bibitem{MishaThese}
  M.~Mikhasenko [COMPASS Collaboration], 
  ``EPJ Web of Conferences'', these proceedings.


\end{thebibliography}
\end{document}